\documentclass[preprint]{vgtc}               

\graphicspath{{figures/}{pictures/}{images/}{./}} 

\usepackage{times}                     

\usepackage{tabu}                      
\usepackage{booktabs}                  
\usepackage{lipsum}                    
\usepackage{mwe}                       

\usepackage{mathptmx}                  

\preprinttext{To appear in the 3rd Workshop on Accessible Data Visualization at IEEE VIS 2026.}

\usepackage{balance} 
\usepackage[hyphens]{url}
\usepackage{breakurl}

\vgtcinsertpkg

\newcommand{\iDotE}{i.\,e.,\xspace}
\newcommand{\eDotG}{e.\,g.,\xspace}

\title{Reflections on Working with Older Adults in Visualization Research}

\author{Zack While\thanks{e-mail: zwhile@ysu.edu}}
\affiliation{\scriptsize Youngstown State University}

\abstract{
    While older adults represent a growing proportion of the global population, their presence in visualization research remains limited. In this paper, I present reflections from a series of human-subject studies conducted with older adults as part of a multi-year research effort. These studies include a controlled laboratory experiment, online evaluations, and an in-situ investigation with participants above age 60. Based on these experiences, I provide methodological takeaways for conducting visualization research with older participants and propose directions for future work. This work ultimately aims to provide practical guidance and encourage broader inclusion of older adults as participants in visualization research.
}

\begin{document}

\firstsection{Introduction}

\maketitle

\noindent 
Advancements in living standards and increased life expectancy are driving a global demographic shift toward an older population~\cite{Ageingan92:online, unWorldPopulation} that is also staying in the workforce longer~\cite{Numberof42:online}. This has coincided with the development of technologies that can empower older adults and enhance their quality of life~\cite{mannheim2019inclusion} through medicine~\cite{le-2016-eval}, physical activity~\cite{fanning2018mobile}, and finances~\cite{price2016effects}; however, poor design can inadvertently hinder their adoption~\cite{chung2023community}. The visualization community has recently pushed for research to focus on broader audiences~\cite{lee2020reaching} while HCI emphasized greater focus on the world's aging population~\cite{lazar2025hci}. Consequently, there is growing attention to understanding how visualization can better support older adults~\cite{backonja2016visualization,while2024gerontovis}. As this area of research grows, however, there is a corresponding need for methodological guidance, as established research practices may not always account for the needs and experiences of aging populations.

Motivated by this need, my dissertation investigated how visualization research can better support older adults through a series of empirical studies, an area we named \textit{GerontoVis}~\cite{while2024gerontovis}. Beyond generating design insights, this work revealed recurring methodological considerations when designing and conducting studies with older adults. This paper thus utilizes that experience to provide a methodological reflection on conducting visualization research with older adults. Specifically, I synthesize lessons learned from designing and conducting multiple human-subject studies with this group, providing guidance and encouraging further research.

\section{Related Work}\label{sec:background}
This section provides necessary background to understand existing research focused at the intersection of aging and data visualization.  Then, to contextualize the lessons described in this paper, I provide brief overviews of the studies that built up my experience.

\subsection{Broader Work in GerontoVis}
Visualization research has examined how aging affects perception and cognition in areas such as working memory~\cite{li2026visual}, cognitive load~\cite{Le2012-integrated,morey-heart-failure}, and spatial processing~\cite{le2014elementary,shao2026experimental}. These changes have motivated a wide range of design considerations, supporting older adults in interpreting and interacting with data.  As a result, design considerations such as visualization selection~\cite{jones-2012,kubota2026practical}, narrative~\cite{errey2026age}, and scaffolding for interpretation~\cite{le2015evaluation,tao2018} have been considered with older adults in mind.  At the same time, older adults have been observed to benefit from accumulated or \textit{crystallized} knowledge~\cite{ursula1989aging}, which in a visualization context may manifest as visualization literacy~\cite{hakone2016} as well as factors regarding designs that are familiar~\cite{doyle-2015,Reeder2014-gk} or grounded in lived experiences~\cite{ahmed2019visualization}.  Beyond visual encodings themselves, the ability to interpret data is also shaped by contextual information.  Context for presented information is also important for proper interpretation~\cite{ahmed2019visualization,alexander-passive}, which, when absent, can lead to reducing a tool's perceived trustworthiness~\cite{le-2018-understanding} or usefulness~\cite{doyle-2015}.  Specific application areas such as patient-provider communication~\cite{reeder2013}, decision-making~\cite{price2016effects}, and behavioral change~\cite{bollen2026bloom} have also received some research attention. 

Despite this body of work, many of these findings are fragmented across specific domains, authors, and design considerations~\cite{while2024gerontovis}, highlighting the need for broader methodological reflection.

\begin{table*}[h]
\centering
\caption{Key papers informing this work, as well as their number of older adult participants, minimum participant age, maximum participant age, and modality.  Note that two papers collected age data via ranges, so the maximum participant age cannot be determined\cite{while2024dark,while2025toward}.}
\label{tbl:key-papers}
\begin{tabular}{lclcll}
\hline
\multicolumn{1}{c}{\textbf{Paper Title (Shortened)}} & \textbf{\# Older Adults} & \textbf{Min Age} & \textbf{Max Age} & \multicolumn{1}{c}{\textbf{Modality}} & \multicolumn{1}{c}{\textbf{Study Type}}\\ \hline
\textit{Glanceable Data Visualizations for Older Adults...}\cite{while2024glanceable} & 24 & 65 & 96 & In-Person & Perceptual Study \\
\textit{Dark Mode or Light Mode?...}\cite{while2024dark} & 66 & 60 & Unknown & Online & Perceptual Study\\
\textit{Toward Filling a Critical Knowledge Gap...}\cite{while2025toward} & 148 & 60 & $ > 75$ & Online & Perceptual Study\\
\textit{Toward Understanding the Experiences of People in...}\cite{while2024toward} & 11 & 76 & 94 & In-Person & Focus Group\\ \hline
\end{tabular}
\end{table*}

\subsection{Published Works Informing This Paper}
My first study in this area was motivated by older adults' increasing usage of wearable devices such as smart watches to track and view their own data~\cite{backonja2016visualization}.  As such, we executed a replication study of Blascheck et al.~\cite{blascheck2018glanceable} examining the impact of aging on the time needed to view a smartwatch visualization and perform a simple data comparison~\cite{while2024glanceable}.  In the study, we investigated three visualization types (\textit{bar}, \textit{donut}, and \textit{radial}) depicting three different amounts of data (7, 12, and 24 data points), splitting the older group into \textit{young-old}~(65-74, $N=12$) and \textit{old-old}~(75 and older, $N=12$) groups.  We found that older participants encountered broad difficulty with the radial chart and benefited more from donut charts than younger participants. Furthermore, we found strong evidence of differences in time thresholds between the young-old and old-old groups.  My work then shifted focus to a non-data encoding component of visualization design, investigating the impact of an interface's \textit{contrast polarity} (\iDotE light and dark mode) on performance (speed and accuracy) for both younger and older adults~\cite{while2024dark}.  While often framed as an accessibility feature, it was unclear whether this was also the case for data visualization.  This suggested that aging does not make contrast polarity more impactful on performance, and that contrast polarity impacted speed as much as visualization type.  We also found that, in both groups, a majority of participants' best-performing polarity did not match their preferred one.

After that, we aimed to dig deeper into understanding how age can affect visual analysis by replicating a study by Saket et al.~\cite{saket2018task}, examining the interactions of aging with low-level visualization tasks (\eDotG finding extreme values and retrieving values) and visualization type (bar charts, line charts, pie charts, scatterplots, and tables)~\cite{while2025toward}.  Using Bayesian modeling to tease out both population- and individual-level differences, we observed that older participants generally required more time to complete the tasks, however accuracy between the two age groups was noticeably similar.  Furthermore, the best-performing visualization types differed between the age groups, with greater variance exhibited by older adults.  Finally, we turned our attention to information displays that older adults use in their daily life. Embedded information displays (EIDs) are interfaces that present information relevant to a device's operations, now common in home appliances like ovens, washers, and thermostats that may pose usability issues for older adults due to age-related changes~\cite{ ghorayeb2023development,tsuchiya2018study} via design concerns also noted in visualization studies~\cite{le-2016-eval,morey2019mobile}. We thus conducted a set of focus groups to qualitatively analyze older adults experiences with these displays at a nearby retirement home~\cite{while2024toward}.  The thematic analysis indicated the existence of several compensatory strategies aimed at circumventing difficulties interacting with and reading from their devices.  Additional information about all four papers is provided in~\autoref{tbl:key-papers}.

\section{Lessons Learned}\label{sec:lessons}
Here I discuss major learning outcomes from completing a dissertation focused on older adults as a target audience.

\subsection{Heterogeneity and Individual Differences}
During the analysis for the task--visualization study~\cite{while2025toward}, we observed several instances in which older participants appeared more heterogeneous than younger ones. This was especially evident when examining the tasks where participants achieved their best performance. Whereas more than 85\% of younger adults shared the same sets of top-, middle-, and bottom-ranked tasks, the corresponding agreement among older adults was substantially lower, with the highest shared proportion reaching only 67\%. This difference was particularly apparent for the middle-ranked tasks, where the five most-common groups of tasks among older adults were shared by 10\% (twice), 11\%, 15\%, and 35\% of participants.

A similar pattern emerged in the glanceable visualizations study~\cite{while2024glanceable}. In several conditions, the confidence intervals associated with the \textit{older} group were noticeably wider than those of the younger group. Similar comparisons between the \textit{young-old} (ages 65-74) and \textit{old-old} ($\geq$ 75) subgroups even suggested variability within the older participants. Thus, researchers should avoid treating older adults as a homogeneous group, as doing so may obscure important individual qualities that impact performance. When sample sizes permit, I recommend reporting and analyzing results separately for commonly used age groups such as the \textit{young-old} and \textit{old-old}, as these groups may reveal differences that are obscured when older adults are analyzed as a single category.  Collecting raw age values rather than age ranges can further support this goal by providing a more granular understanding of age-related diversity and enabling analyses beyond predefined age categories.  However, age groups alone may not fully capture this diversity, and attributes such as technology experience, health, and disability status can sometimes be more informative when explaining observed differences.  When recruiting participants, researchers should also aim for representation from a wide variety of ages above $60$ whenever possible and report their relative proportions.  This aligns with recent work spotlighting the study of individual differences~\cite{liu2020survey} and could motivate the use of data analysis methods that effectively model them, such as Bayesian hierarchical modeling~\cite{davis2022risks}.

\subsection{Participant Recruitment}
Participant recruitment for in-person studies involving older adults presents unique challenges, as has been noted in broader aging research~\cite{forsat2020recruitment,mchenry2015recruitment,mody2008recruitment} as well as in human subjects research for accessibility~\cite{mack2022anticipate}. Based on my experiences, I highlight several practical considerations for researchers working with this population. For the glanceable study~\cite{while2024glanceable}, recruitment involved identifying community bulletins targeted toward older adults, both digital and in-person. While both were useful, in-person bulletin boards appeared to be significantly more effective, suggesting that physical community spaces may remain valuable recruitment channels for this population.  The in-person bulletin boards may have been more effective based on their proximity to where older adults in the area gathered (\eDotG a community center), so it is important to consider advertisements locations.  In contrast, the EID study~\cite{while2024toward} relied on collaboration with a nearby retirement home, where recruitment required interacting with both staff and residents. Providing a short introductory presentation proved particularly beneficial, allowing for direct engagement with potential participants, clarification of study goals, and informal snowball sampling. I thus highly recommend this approach to proactive communication and relationship-building.

Additionally, logistical and communication preferences may differ for older adults recruited for in-person studies. Participants for those studies often preferred phone calls for initial discussions to clarify details and emails for scheduling, and using a local phone number helped alleviate concerns about unfamiliar contact information; this aligns with the \textit{Communication} dimension for recruitment described by Mack et al.~\cite{mack2022anticipate}.  Multiple participants required additional scheduling flexibility due to having a family member or caretaker driving them to and from the study.  Ultimately, successful recruitment with older adults seems to benefit from creating in-person connections with the community, establishing trust through direct interaction, and adapting communication methods to participant preferences.  However, this work occurred in the U.S., so some takeaways may not apply broadly to other areas and cultures. 

\subsection{Task Completion Time Across Age Groups}
Both the glanceable visualizations study~\cite{while2024glanceable} and the task--visualization study~\cite{while2025toward} demonstrated an interesting observation in slightly different ways: older adults can achieve approximately the same accuracy as younger adults while performing visual analysis tasks, however they appear to require more time to do so. This pattern appeared across several tasks, multiple visualization types, and two experimental settings, suggesting that aging may influence the efficiency of visual analysis rather than its correctness.  This finding both conflicted with some existing results~\cite{liu2022contextualizing} and challenged assumptions that aging necessarily affects performance~\cite{le-2016-eval}, instead pointing toward a more nuanced relationship between aging and visual analysis.  As such, time becomes a critical, underexamined performance metric, raising methodological questions regarding study duration, pacing, and the interpretation of performance differences across age groups. Researchers should therefore collect and report completion times alongside accuracy whenever possible, rather than treating accuracy as the sole measure of performance. Studies should also account for potentially longer completion times by providing flexible pacing and allocating additional session time for older participants.  When study duration becomes a concern, researchers may also need to reduce the number of tasks or conditions to avoid participant fatigue while still allowing sufficient time for task completion.  Broader accessibility work has similarly emphasized flexible pacing, breaks, and accommodations for participants with varying abilities and fatigue levels~\cite{mack2022anticipate,mack2022chronically}. This motivates further work investigating the nuanced impact of aging on task speed.

\subsection{Online and In-Person Studies Offer Complementary Insights}
\begin{figure}
    \centering
    \includegraphics[width=0.5\linewidth]{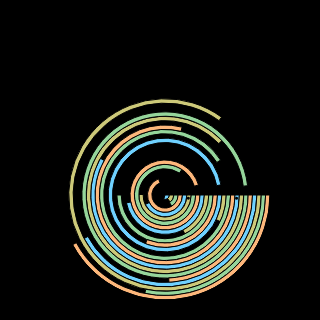}
    \caption{An example of the \textit{radial-24} condition from the glanceable study~\cite{while2024glanceable}, which required identifying two of the twenty-four bars via a black dot and selecting the one (left or right) that represented a larger value.  Initial participants described difficulty focusing on several of these visualization on a 1.6-inch screen for extended periods of time, resulting in a modification to the procedure for them to skip it.  The image shown is used with permission~\cite{blascheck2018glanceable}.}
    \label{fig:radial24}
\end{figure}

\begin{figure}
    \centering
    \includegraphics[width=\linewidth]{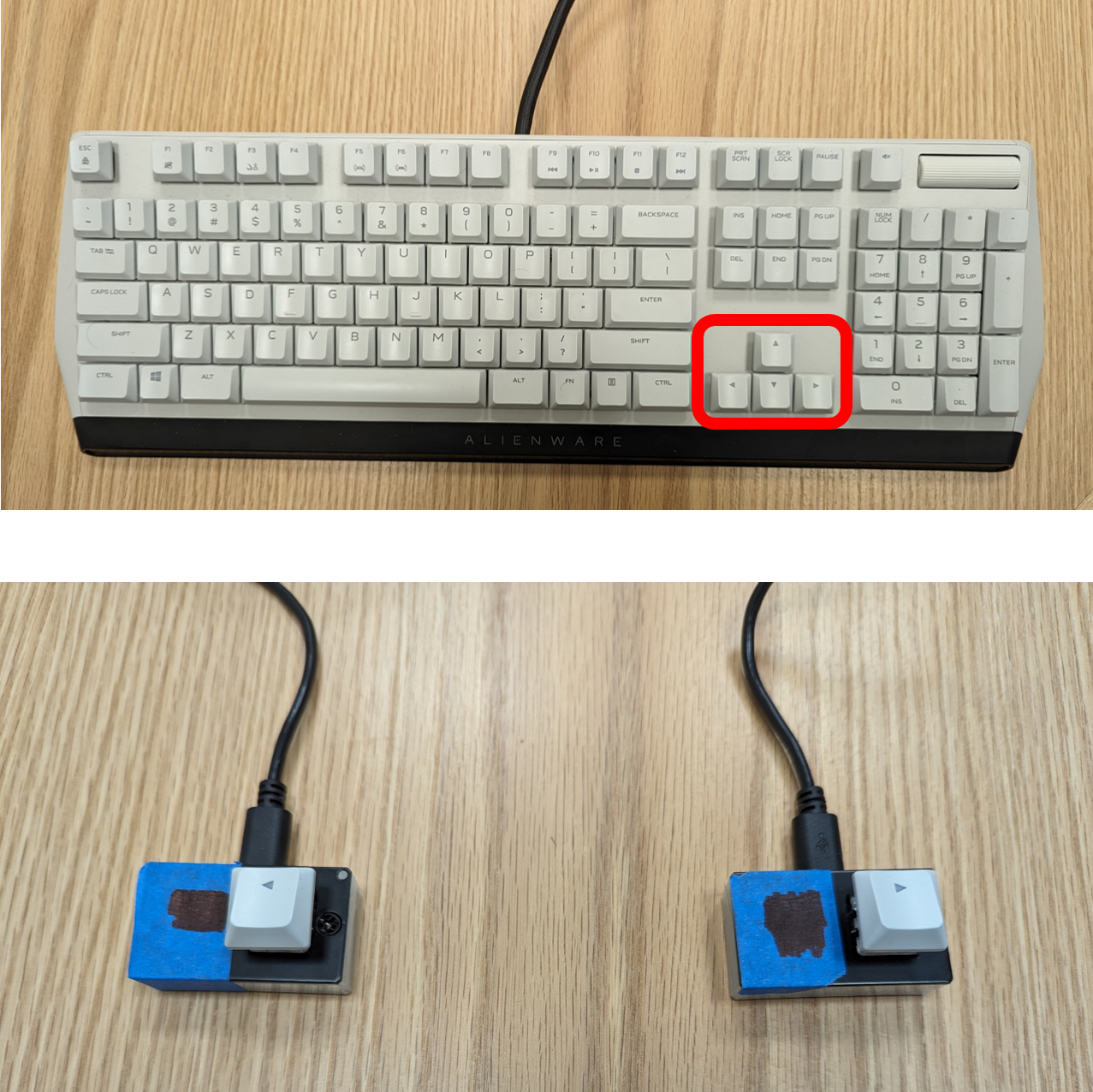}
    \caption{The original (top) participant input device from the glanceable study~\cite{while2024glanceable} used the arrow keys, which caused participant discomfort.  This was replaced by two separate keys (bottom), which gave participants the freedom to place them in a more comfortable position.  Images are being reused with permission~\cite{while2024glanceable}.}
    \label{fig:devices}
\end{figure}

The EID~\cite{while2024toward} and glanceable visualization~\cite{while2024glanceable} studies were conducted in-person, whereas the contrast polarity~\cite{while2024dark} and task--visualization~\cite{while2025toward} studies were conducted online using Prolific~\cite{Prolific80:online}. While both modalities proved valuable, they often exposed different aspects of the older adult experience.  Conducting studies in-person provided contextual insights that would have been difficult to obtain remotely. For example, touring apartments and shared spaces within the retirement community before the EID focus groups improved the specificity of discussion prompts and revealed technologies and appliances commonly used by residents. Likewise, conducting the glanceable study in-person caused us to observe participant discomfort looking at one type of visualization for several minutes (shown in~\autoref{fig:radial24}), which required modifying the procedure~\cite{while2024glanceable}.  Furthermore, the study's primary input peripheral~(shown in~\autoref{fig:devices}), was causing physical discomfort due to being pressed repeatedly for the entire study (45-75 minutes), requiring us to quickly find a more ergonomic replacement~\cite{while2024glanceable}.

In contrast, online studies highlighted that participants recruited through Prolific were often highly computer literate and comfortable using technology. While this enabled efficient recruitment of older adults at scale, it also raised questions about how representative such participants were of the broader aging population. Consequently, researchers should not assume that older adults recruited online are representative of older adults more broadly. When possible, I recommend reporting participants' technology experience, recruiting through multiple channels (\eDotG online platforms and community organizations), and explicitly acknowledging which segments of the older adult population may be underrepresented.  At the same time, online studies provide valuable access to older adults who are comfortable using technology, making them well-suited for rapidly collecting larger samples and studying technology-oriented populations.  Online studies are also well-suited for any type of study where personal or home items are of interest (\eDotG co-design studies~\cite{cajamarca2022codesign}) or transportation for a participant may be difficult~\cite{mack2022anticipate}, since the study can take place via a video call.  For research involving technology adoption, accessibility, or everyday use cases, supplementing online recruitment with at least a small in-person version of a study may help reveal otherwise-hidden challenges.

\subsection{Considerations for Replication}
Both the glanceable~\cite{while2024glanceable} and task--visualization studies~\cite{while2025toward} were designed as \textit{conceptual replications} of prior visualization research~(\iDotE~\cite{blascheck2018glanceable,saket2018task}), examining whether findings established with younger participants would generalize to older adults. To accomplish this, we recruited only older adults and compared their performance against the younger cohorts reported in the original studies.  While this approach was efficient, it introduced differences in experimental setup, recruitment procedures, and study context that complicated age-based comparisons. In retrospect, recruiting younger and older adults within the same study would have provided a stronger basis for evaluating age-related differences while simultaneously testing the generalizability of prior findings. Therefore, when adapting existing visualization studies for older adults, I recommend recruiting both age groups whenever feasible. This allows observed differences to be more confidently attributed to age rather than methodological variation between the two studies.

\subsection{Participant Remuneration}

Participant remuneration also presented unexpected situations, particularly for in-person studies with older adults. Our initial plan was to offer Amazon gift cards distributed digitally via email; however, several participants either did not use Amazon or expressed discomfort with online shopping, citing reasons such as lack of familiarity, distrust, or personal preferences. As a result, some participants declined compensation or attempted to transfer it to others.  To better align with participant needs, we transitioned to offering physical gift cards sent via mail, with a broader selection of stores (\eDotG Wal-Mart). This approach reduced confusion and improved overall acceptance of compensation. With this in mind, I warn that remuneration strategies commonly used in online studies or with younger populations may not necessarily generalize to older ones, and that offering flexible, tangible, and widely accessible compensation options is important when working with older populations.
\section{Future Work}
This work also highlights several directions for future methodological development in GerontoVis. A key limitation of many of the studies presented here is that they were conducted either online or within the \textit{Five Colleges} region of Amherst, Massachusetts, resulting in samples that tended toward higher education levels and greater technology and visualization familiarity. While this did not appear to substantially impact the observed results, it suggests a need for future studies that examine how well existing visualization methodologies generalize to older adults with more diverse backgrounds and life experiences. In particular, future work should explore recruitment and engagement approaches (\eDotG~\cite{peck2019data}) that better support participation from underrepresented populations.

Additionally, this work has primarily focused on accuracy and time as measures of visualization effectiveness.  However, these are not the only relevant metrics, and future work should investigate whether the evaluation approaches commonly used in visualization research adequately capture older adults' experiences. Factors such as memorability, attention, trust, confidence, cognitive workload, and interaction burden may be equally important when assessing visualization use among aging populations. Furthermore, GerontoVis would benefit from methodological examinations of whether existing evaluation techniques are sufficient or whether new measures and study protocols are needed to better characterize how older adults interact with and benefit from data visualizations.
\section{Conclusion}
In this paper, I reflected on lessons that I learned after completing a dissertation focused on conducting data visualization human subjects studies with older adults.  This included takeaways from empirical results, logistical and methodological observations about the process that I wish I had known beforehand, and ideas for future work in this research area.  I ultimately hope that these learning outcomes can assist future researchers interested in GerontoVis.

\balance
\bibliographystyle{abbrv-doi}
\bibliography{bibliography}

\end{document}